\documentclass[aps,prb,twocolumn,groupedaddress,superscriptaddress,nofootinbib,notitlepage,showpacs,floatfix]{revtex4-1}
\usepackage{graphicx,graphics,epsfig,times,bm,bbm,amssymb,amsmath,amsfonts,amsthm,mathrsfs,multirow}
\usepackage[pdfstartview=FitH]{hyperref}
\hypersetup{
colorlinks=true,       	
linkcolor=blue,          	
citecolor=blue,        
filecolor=magenta,      	
urlcolor=blue,           	
runcolor=cyan
}
\newcommand{\ignore}[1]{}
\usepackage{epstopdf}
\usepackage{dcolumn}
\usepackage[pdftex]{color}
\usepackage{braket}
\usepackage{enumerate}
\begin{document}

\title{
% [A]
Symmetry fractionalization: \\
Symmetry-protected topological phases of the bond-alternating spin-$1/2$ Heisenberg chain}

\author{R. Haghshenas}
\affiliation{Department of Physics, Sharif University of Technology, P.O.Box
11155-9161, Tehran, Iran}
%\affiliation{Department of Physics, Sharif University of Technology, Tehran 14588, Iran}
\email{haghshenas@physics.sharif.edu}

\author{A. Langari}
\affiliation{Department of Physics, Sharif University of Technology, P.O.Box
11155-9161, Tehran, Iran}
%\affiliation{Department of Physics, Sharif University of Technology, Tehran 14588, Iran}
\affiliation{Center of Excellence in Complex Systems and Condensed Matter, Sharif
University of Technology, Tehran 1458889694, Iran}
\affiliation{Max-Planck-Institut f\"ur Physik komplexer Systeme, 01187 Dresden, Germany}

\author{A. T. Rezakhani}
\affiliation{Department of Physics, Sharif University of Technology, P.O.Box
11155-9161, Tehran, Iran}
%\affiliation{Department of Physics, Sharif University of Technology, Tehran 14588, Iran}
%%%%%%%%%%%%%%%%%%%%%%%%%%%%%%%%%%%%%%%%%%%%%%%%%%%%

\begin{abstract}
We study different phases of the one-dimensional bond-alternating spin-$1/2$ Heisenberg model by using the symmetry 
fractionalization mechanism. We employ the infinite matrix-product state representation of the ground state (through the infinite-size density matrix renormalization group algorithm) to obtain inequivalent projective representations of the (unbroken) symmetry groups of the model, which are used to identify the different phases. We find that the model exhibits trivial as well as symmetry-protected topological phases. The symmetry-protected topological phases are Haldane phases on even/odd bonds, which are protected by the time-reversal (acting on the spin as $\bm{\sigma}\rightarrow-\bm{\sigma}$), parity (permutation of the chain 
about a specific bond), and dihedral ($\pi$-rotations about a pair of orthogonal axes) symmetries. Additionally, we investigate the phases of the most general two-body bond-alternating spin-$1/2$ model, which respects the time-reversal, parity, and dihedral symmetries, and obtain its corresponding twelve different types of the symmetry-protected topological phases. 
\end{abstract}

\maketitle
%%%%%%%%%%%%%%%%%%%%%%%%%%%%%%%%%%%%%%%%%%%%%%%%%%%%

\section{Introduction} 
\label{sec:intr}

For many years, Landau-Ginzburg theory of phase transitions was the dominant paradigm for characterization of different phases of matter \cite{book:Goldenfeld}. In this theory the characterization is based on the breaking of a symmetry associated with a local order parameter. Over the past decade, however, the emergence of some exotic phases such as the Haldane phase \citep{Haldane-1983-Continuum} in one dimension, $Z_{2}$ spin liquids \cite{Read-1991-Expansion}, topological insulators \citep{Kane-2005-Quantum}, and quantum Hall states \citep{Tsui-1982-Two} (which all elude Landau-Ginzburg theory) has attracted a renewed interest in characterization of quantum phases. In particular, in one dimension, it has been proven that quantum phase transitions between the so-called ``symmetry-protected topological" (SPT) phases are not accompanied by any symmetry breaking\cite{Pollmann-2010-Entanglement}. Thus providing powerful methods for reliable classification of phases is still much needed and of fundamental importance.

Recently, based on symmetries of a given model and corresponding transformation of matrix-product state (MPS) representation of its ground state, a ``symmetry fractionalization" scheme to classify phases has been proposed in Refs.~\onlinecite{Pollmann-2010-Entanglement,Turner-2011}. Later, by combining the symmetry-breaking mechanism of Landau-Ginzburg theory and the symmetry fractionalization technique, a unified formalism for identification of phases of one-dimensional gapped systems was developed \cite{XieChen-2011-Classification, XieChen-2011-Complete}. In one dimension, this picture is complete, but in the case of higher dimensions it is not. In brief, this method employs the projective representations \cite{Projective-note} of the (unbroken) symmetry groups of the underlying model to assign a set of unique \textit{labels} for each phase---for a short review, see Appendix \ref{App}. 

 Finding appropriate order parameters to identify SPT phases has been the subject of vast recent investigations \citep{Cui-2013-Local,Marvian-2013-Symmetry,Haegeman-2012-Order,Zaletel-2013-Detecting,Huang-2013-Detection}. Most of the proposed order parameters cannot determine crucial characteristics of SPT phases, but still may show the presence or absence of a SPT phase. To fully characterize SPT phases, a direct calculation of inequivalent classes of projective representations of the symmetry groups of the system of interest is required. In particular, in Ref.~\onlinecite{Pollmann-2012-Detection} an order parameter based on iMPS representation of ground state has been introduced, which can find fractionalizations of (unbroken) symmetry groups. It has been argued that this order parameter is sufficiently strong in order to specify \emph{any} SPT phases---for another strong order parameter, see Refs.~\onlinecite{Haegeman-2012-Order,Langari-2013-Ground}. 

Here, we employ the iMPS representation of the ground state to study the spin-$1/2$ bond-alternating 
Heisenberg chain and find the associated entanglement spectrum (the eigenvalues of the half-system 
reduced density matrix). This spectrum shows a significant change in the behavior of the model through a 
potential quantum phase transition point, observed in the form of evenness and oddness of the entanglement 
spectrum degeneracy. However, the entanglement spectrum is insufficient to provide a picture in which all 
phases are specified. This goal, in fact, requires the stronger symmetry fractionalization technique. 
We incorporate the known symmetries of the model in the iMPS representation, and calculate its phase labels. 
Specifically, we demonstrate that the model has two SPT phases and a symmetry-broken (ferromagnetic) phase. 
These two SPT phases are due to the existence of two different types of bonds (called ``odd" and ``even", 
or ``red" and ``blue", respectively) corresponding to the two couplings of the model ($J$ and $J'$, respectively). 
The labels of these phases are evidently different; however, part of this label set in each is exactly equal 
to the labels of the Haldane phase (throughout the paper, the Haldane phase is referred to the phase 
characterized by the ground state of the spin-$1$ Heisenberg chain or equivalently to the 
Affleck-Kennedy-Lieb-Tasaki (AKLT) model \cite{Affleck-1987-Rigorous}). 
Next, we study the most general spin-$1/2$ bond-alternating model respecting the time-reversal ($T$), 
parity ($P$), and dihedral ($D_2$) symmetries, and classify twelve different kinds of SPT phases
among several possible phases. 
To see that whether there is any hidden symmetry responsible for the obtained phase diagram, 
we perturb the bond-alternating Hamiltonian with three different symmetry-breaking terms. 
As a result, we conclude that the phase portrait obtained through the $[TPD_2]$ symmetries is already complete.     

This paper is organized as follows. In Sec.~\ref{sec:model} we describe the model and how an iMPS representation for its ground state can be constructed. Next in Sec.~\ref{sec:classification} we identify the associated phases of the model by the symmetry fractionalization. We also use the iMPS representation in order to obtain the entanglement spectrum. To see how general and rich the phase landscape of the model can be, we also investigate the SPT phases of the most general bond-alternating model. In addition, we discuss how the phases are protected by the $P$, $T$ and $D_2$ symmetries. Section~\ref{sec:conc} concludes the paper, and summarizes our main findings. Appendix \ref{App} includes details of the symmetry fractionalization technique and its application to the bond-alternating model.
%%%%%%%%%%%%%%%%%%%%%%%%%%%%%%%%%%%%%%%%%%%%%%%%%%%%
\begin{figure}
\includegraphics[scale=.67]{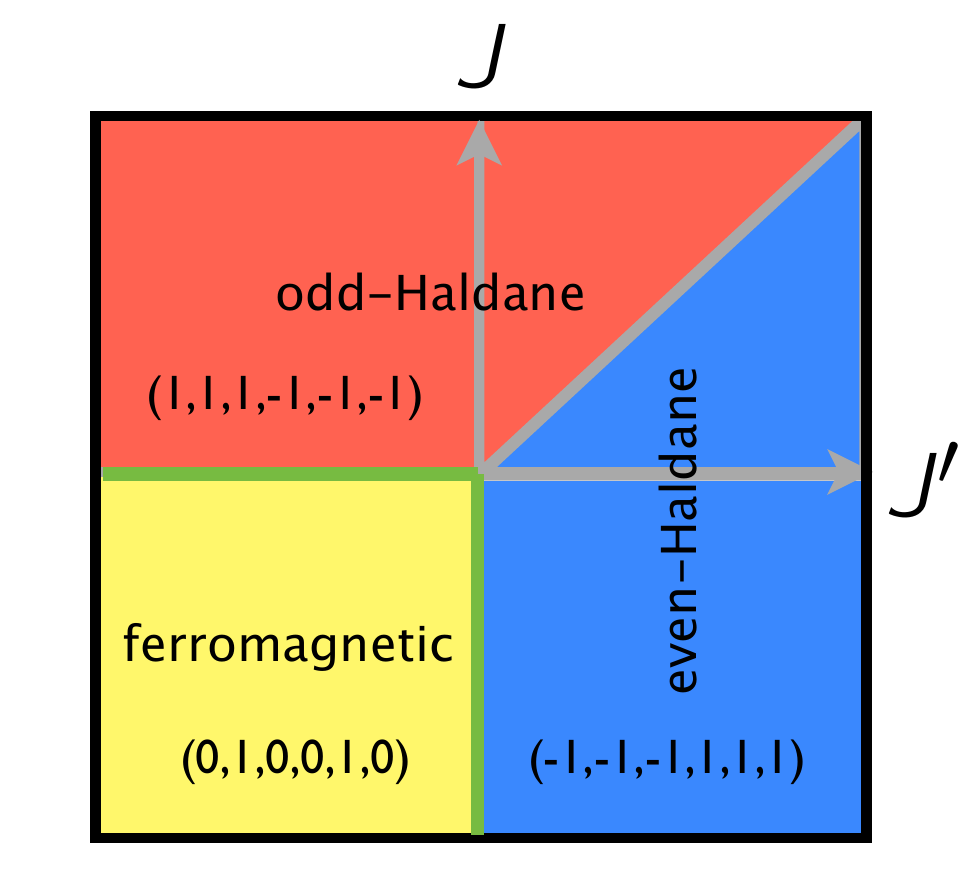}
\caption{(Color online) Phase diagram of the one-dimensional bond-alternating spin-$1/2$ Heisenberg model \cite{Wang-2013-Topological}. Here $J$ and $J'$ are the exchange couplings associated to the even and odd bonds, respectively. In the boundary of the odd- and even-Haldane phases a topological phase transition occurs. The labels $(0,1,0,0,1,0)$, $(1,1,1,-1,-1,-1)$, and $(-1,-1,-1,1,1,1)$ identify the trivial (ferromagnetic), odd-Haldane, and even-Haldane phases, respectively---see subsection~\ref{subsec:sym-frac}.}
\label{fig:phase-diagram}
\end{figure}

%%%%%%%%%%%%%%%%%%%%%%%%%%%%%%%%%%%%%%%%%%%%%%%%%%%%

%%%%%%%%%%%%%%%%%%%%%%%%%%%%%%%%%%%%%%%%%%%%%%%%%%%%
\section{Bond-alternating spin-$1/2$ Heisenberg model}
\label{sec:model}

%%%%%%%%%%%%%%%%%%%%%%%%%%%%%%%%%%%%%%%%%%%%%%%%%%%%
\subsection{Hamiltonian} 

The bond-alternating spin-$1/2$ Heisenberg model on a one-dimensional chain 
is defined by the following Hamiltonian:
\begin{equation}
H= \sum_{i=1}^{\infty} J\bm{\sigma}^{(2i-1)}\cdot\bm{\sigma}^{(2i)}+ J'\bm{\sigma}^{(2i)}\cdot\bm{\sigma}^{(2i+1)},
\label{Hamt}
\end{equation} 
where $\bm{\sigma}=(\sigma_x, \sigma_y, \sigma_z)$ are the Pauli operators, and $J$ and $J'$ are the exchange couplings. We argue later that the ratio of the very couplings controls a topological phase transition. The excitation spectrum of the model and its connection to the spin-Peierls transition has been studied in Ref.~\onlinecite{Bonner-1982-Excitation}, where it has been shown that the gapless phase of the homogeneous spin-$1/2$ Heisenberg chain ($J=J'$) is unstable against an addition of bond-alternation to a gapped spin-Peierls state. Figure \ref{fig:phase-diagram} shows a sketch of the phase diagram of the model, which has been obtained by implementing iMPS with the infinite time-evolving block decimation (iTEBD) method \cite{Wang-2013-Topological}. Suitable string order parameters \cite{Hida-1992} have indicated two distinct phases for the blue and red regions of Fig.~\ref{fig:phase-diagram} separated by a quantum phase transition at the gapless line $J=J'>0$, where the central charge $c\approx 1$ implies a Gaussian transition there. Since the string order parameters have exactly the form of string order parameter of the Haldane phase, the phases were named even- and odd-Haldane phases. 

Nevertheless, it should be noted that string order parameters are not reliable labels for phases, since they are not necessarily stable in the Haldane phase\citep{Pollmann-2010-Entanglement}. In addition, nonzero string order parameters usually show the preservation of the symmetry and surprisingly if we choose them according to the projective representation of the symmetries they become zero (the signature of the phases\citep{Pollmann-2012-Detection}). 

The model also represents a ferromagnetic phase when both $J$ and $J'$ become negative, shown by the yellow region in Fig.~\ref{fig:phase-diagram} separated by the solid-green lines from the even- or odd-Haldane phases. The transition to the ferromagnetic phase is accompanied by the spontaneous breaking of the $\mathrm{SU}(2)$ to $\mathrm{U}(1)$ symmetry, which is defined by the magnetization order parameter.

Therefore, to identify the different phases of the model we need to specify the symmetry fractionalization of the model. To this end, we obtain inequivalent projective representations of the symmetry groups of the Hamiltonian (\ref{Hamt}), which include $T$, $P$, $\mathrm{SU}(2)$ rotations, and the two-site translational invariance ($TI$).

\textit{Remark.}---It has been known that spin-$1/2$ models with the $T$ and one-site $TI$ symmetries cannot have any SPT phase; they only show symmetry breaking (degenerate ground states) or gapless phases \cite{XieChen-2011-Classification}.
Thus to observe the SPT phases in the spin-$1/2$ models we need to explicitly break the $TI$ symmetry (as occurred in Eq. (\ref{Hamt})) or use three-body interactions, e.g., as appeared in cluster Hamiltonians \cite{Montes-2012-Phase}.

%%%%%%%%%%%%%%%%%%%%%%%%%%%%%%%%%%%%%%%%%%%%%%%%%%%%
\begin{figure}
\includegraphics[width=7cm,height=4.7cm]{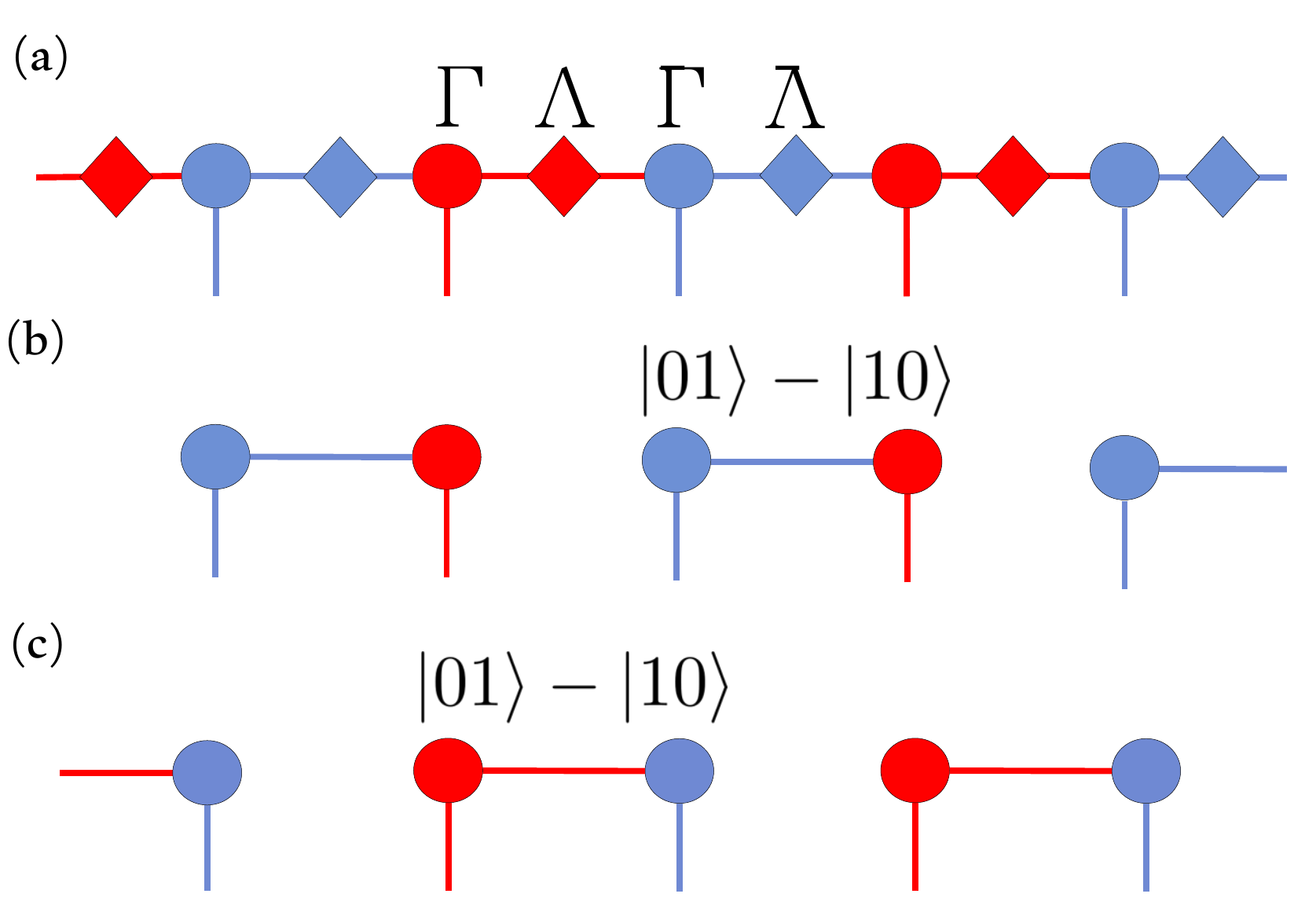}
\caption{ (Color online) (a) Schematic representation of iMPS. Here the blue (red) diamonds indicate $\Lambda$ ($\overline{\Lambda}$), 
where $\varrho_{\mathrm{b}}=\Lambda$$\Lambda^{\dagger}$ ($\varrho_{\mathrm{r}}=\overline{\Lambda}$ $\overline{\Lambda}^{\dagger}$) is the density matrix of the semi-infinite chain, when the chain is partitioned into two parts from the blue (red) bonds. (b) and (c) represent a schematic representation of the ground states of the bond-alternating spin-$1/2$ Heisenberg model when ($J'=0$, $J>0$) and ($J=0$, $J'>0$), respectively.}
\label{fig:MPS}
\end{figure}
   
%%%%%%%%%%%%%%%%%%%%%%%%%%%%%%%%%%%%%%%%%%%%%%%%%%%%
\begin{figure*}[tp]
\vskip5mm
\includegraphics[width=8.5cm,height=5.cm]{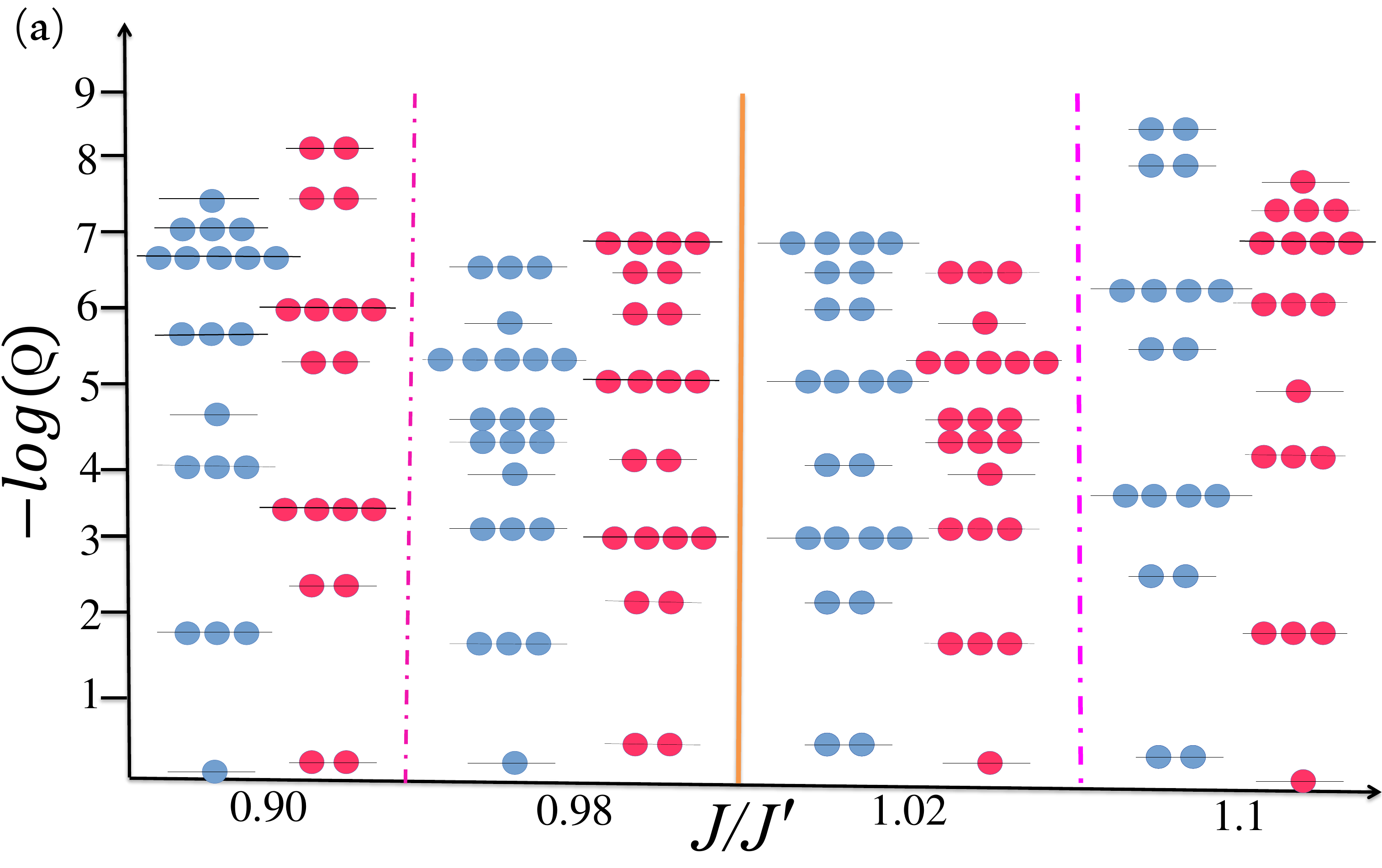}
\hspace{5mm}
\includegraphics[width=8.5cm,height=5cm]{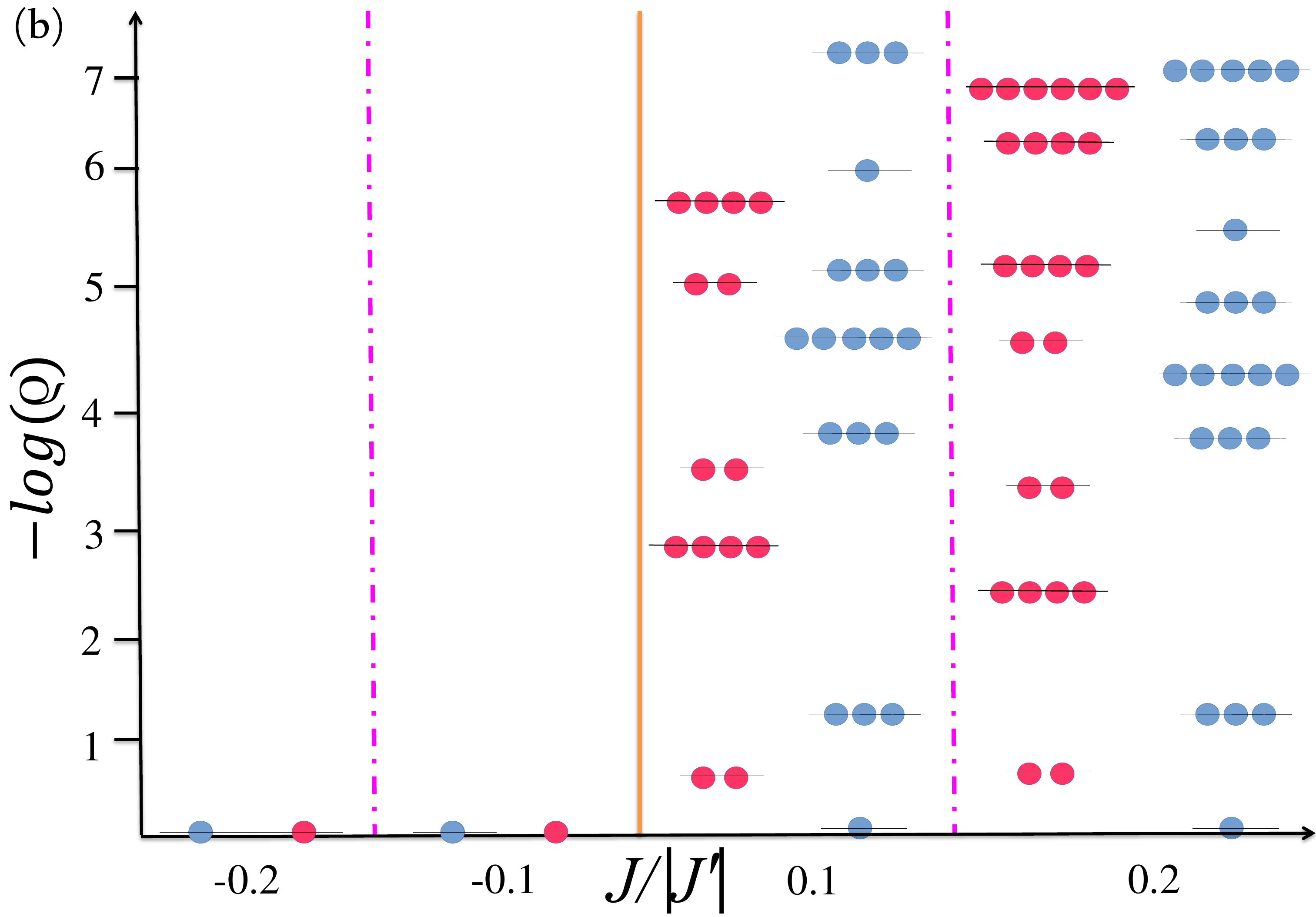}
\caption{(Color online) Entanglement spectrum of the bond-alternating Heisenberg model (\ref{Hamt}).
Vertical axis represents logarithm of the largest eigenvalues of
density matrices $\varrho_{\mathrm{b}}$ (blue points) and $\varrho_{\mathrm{r}}$ (red points).  
(a) The orang solid line is the location  
of quantum phase transition from the even-Haldane phase to the odd-Haldane phase. (b) The degeneracy of the entanglement spectrum suddenly disappears when the system passes through the critical point $(J',J)=(-1,0)$ (orang solid line).} 
\label{fig:Haldane}
\end{figure*}
%%%%%%%%%%%%%%%%%%%%%%%%%%%%%%%%%%%%%%%%%%%%%%%%%%%%

%%%%%%%%%%%%%%%%%%%%%%%%%%%%%%%%%%%%%%%%%%%%%%%%%%%%
\subsection{iMPS representation of the ground state} 
\label{subsec:imps}

We use an iMPS representation scenario to obtain the ground state of the Hamiltonian (\ref{Hamt}). This type of representation has been proven reliable for the ground state of one-dimensional gapped Hamiltonians \cite{Hastings-2007-Area}. To minimize the energy, there exist several algorithms (all resulting an iMPS representation) such as iTEBD \cite{Vidal-2006-Classical}, infinite-size density matrix renormalization group (iDMRG) \cite{Mcculloch-2008-Infinite, Kjall-2013-phase, Schollwock-2011-Density}, and matrix-product operator representations \cite{Pirvu-2010-Matrix, Liu-2010-Symmetry}. Although iTEBD is the dominant method used to enhance the convergence of the algorithm, because of some valuable features of iDMRG, especially fast convergence and no need to apply the Trotter-Suzuki approximation, we adopt iDMRG here (as outlined in Ref.~\onlinecite{Mcculloch-2008-Infinite}). According to the iDMRG algorithm, the ground state of the $N$-site system is given by
\begin{equation}
|\Psi \rangle=\sum\limits_{
m_1,\ldots,m_N
} \mathrm{Tr} [\Gamma_{m_{1}}\Lambda \dots \Gamma_{m_{N}}\Lambda]|m_{1}\dots m_{N}\rangle,
\label{State}
\end{equation}
where $\Lambda$s are some diagonal positive matrices, and $\Gamma_{m_{\ell}}$s are some $\chi\times\chi$ matrices associated to site $\ell$. It is evident that the Hamiltonian (\ref{Hamt}) and its corresponding ground state have the two-site $TI$ symmetry, whereby two pairs $(\Lambda, \overline{\Lambda})$ and $(\Gamma,\overline{\Gamma})$ are needed within the iMPS representation of the ground state. These symbols are shown by blue and red colors in Fig.~\ref{fig:MPS}-(a), respectively. 
We also assume that the two-site transfer matrix $\widehat{T}_{\alpha \beta;ij}\equiv \sum_{mn} (\overline{\Gamma}_{m} \overline{\Lambda} \Gamma_{n} \Lambda)_{\alpha i} (\overline{\Gamma}_{m} \overline{\Lambda} \Gamma_{n} \Lambda)^{\ast}_{\beta j}$ satisfies the following conditions: (i) $\sum_{ij}\widehat{T}_{\alpha \beta;ij}\delta_{ij}= \eta \delta_{\alpha \beta}$, and (ii) the eigenvalue $\eta$ is non-degenerate and maximum---see Appendix \ref{App} and Fig.~\ref{fig:SystemFactor}-(a). 
In this case, the iMPS representation of the ground state is called the ``canonical short-range correlated form." In the canonical short-range correlated iMPSs, $(\Lambda,\overline{\Lambda})$ are the diagonal positive matrices which represent the density matrices of the semi-infinite chain. Note that $\varrho_{\mathrm{b}}=\Lambda \Lambda^{\dagger}$ ($\varrho_{\mathrm{r}}=\overline{\Lambda}$ $\overline{\Lambda}^{\dagger}$) shows the density matrix when the chain is partitioned into two parts from the blue (red) bonds of the iMPS representation of Fig.~\ref{fig:MPS}-(a).

%%%%%%%%%%%%%%%%%%%%%%%%%%%%%%%%%%%%%%%%%%%%%%%%%%%%
\section{Classification of phases}
\label{sec:classification}

%%%%%%%%%%%%%%%%%%%%%%%%%%%%%%%%%%%%%%%%%%%%%%%%%%%%
\subsection{Entanglement spectrum} 
\label{subsec:entanglement-spectrum}

The entanglement spectrum has proven to be a proper candidate to identify the SPT phases without prior knowledge of the symmetry, since the different projective representations of the symmetry manifest themselves in the degeneracy of entanglement spectrum pattern. For example, the presence of the SPT phases results in an even degeneracy of the entanglement spectrum \cite{Pollmann-2010-Entanglement}. Figure~\ref{fig:Haldane}-(a) shows the entanglement spectrum of $\varrho_{\mathrm{b}}$ and $\varrho_{\mathrm{r}}$ vs $J/J'$. In the $J<J'=1$ region, the degeneracies of the eigenvalues of $\varrho_{\mathrm{b}}$ and $\varrho_{\mathrm{r}}$ are, respectively, odd and even, and remain unchanged throughout this region. Here one can conclude that the entanglement of the bonds which possess even degeneracy cannot be removed by using any local unitary transformation unless the system undergoes a phase transition. In contrast, one can adiabatically transform the bonds possessing odd degeneracy to a product state (trivial phase). Right after crossing $J/J'=1$, the degeneracies on the bonds change to even (for blue) and odd (for red), which signals a quantum phase transition---since the change in degeneracies can be associated to a quantum phase transition. 

We now label the phases with the even-Haldane ($J/J'<1$) and odd-Haldane ($J/J'>1$), which will be clearer in the next subsection. Another phase transition can be characterized for $J, J'<0$, where the odd/even degeneracy of the entanglement spectrum disappears, and the entanglement spectrum indicates a dominant single eigenvalue (called the ferromagnetic phase), as depicted in Fig.~\ref{fig:Haldane}-(b). This pattern illustrates either a trivial or a symmetry broken phase. In fact, in the ferromagnetic phase the $T$ and $D_{2}$ symmetries are broken, while $P$ is preserved.

Although the entanglement spectrum reliably signals the SPT phase transitions here, it is in general insufficient to characterize the SPT phases.  Therefore, we implement the symmetry fractionalization to identify the SPT phases.

 %%%%%%%%%%%%%%%%%%%%%%%%%%%%%%%%%%%%%%%%%%%%%%%%%%%%
\subsection{Symmetry fractionalization } 
\label{subsec:sym-frac}

Here we use the symmetry fractionalization mechanism to obtain unique labels for the phases of the bond-alternating spin-$1/2$ Heisenberg chain. We employ the procedure of Appendix~\ref{App} to calculate numerically inequivalent projective representations of the symmetries of the model, which include $T$, $P$, $\mathrm{SU}(2)$, and two-site $TI$. We then compare the labels of the phases of the model with the Haldane phase.   

Due to the two-site $TI$ symmetry, unitary operations on two-site blocks determine inequivalent projective representations of symmetries. In this case, inequivalent projective representations of $\mathrm{SU}(2)$ in the presence of two-site $TI$ are specified by the $D_{2}$ with two-site $TI$ symmetries. If we denote $u(x)=\sigma_{x}$ and $u(z)=\sigma_{z}$, then the representation of the symmetry group $D_{2}$ is defined by $G_{D_{2}}=\langle u(x)\otimes u(x),u(z)\otimes u(z)\rangle$. Here $\langle\cdots\rangle$ denotes the generators of the group, which satisfy $u(x)u(z)=u(z)u(x)$ and $u(x)u(x)=u(z)u(z)=\openone$. To preserve the $D_{2}$ symmetry in the iMPS representation of the ground state with the two-site $TI$, $\Gamma$ and $\overline{\Gamma}$ are required to satisfy
\begin{align}
\sum_{i} u_{ji}(g)\Gamma_{i}&=\beta(g) Z^{-1}(g)\Gamma_{j}X(g), \label{SUO}\\
\sum_{i} u_{ji}(g)\overline{\Gamma}_{i}&=\beta'(g) X^{-1}(g)\overline{\Gamma}_{j}Z(g),
\label{SU2O}
\end{align} 
where $\beta(g)$ and $\beta'(g)$ are arbitrary phases, and the $g$ index ($g\in \{x,z\}$) represents the elements of the $D_{2}$ symmetry. The inequivalent projective representations of $D_{2}$ and two-site $TI$ are given by $X(x)X(z)=\pm X(z)X(x)$ and $Z(x)Z(z)=\pm Z(z)Z(x)$, which introduce four different classes of projective representation of the symmetries. Within each specific phase, one of the above projective representations holds, and it can change to the other ones only through a quantum phase transition. Therefore, the $\pm$ signs actually provide  a unique label for the phases. To determine the projective representation (within each phase), we need to calculate the corresponding $\pm$ signs of the projective representations. Accordingly, we define the parameters
\begin{align*}
\Pi^{\mathrm{b}}_{D_{2}}&=
\left\{
  \begin{array}{cl}
0&\text{if}\ \ \ |\eta'|<1\\
(1/\chi)\mathrm{Tr}[{X(x)X(z)X^{-1}(x)X^{-1}(z)}]&\text{if}\ \ \ |\eta'|=1
  \end{array}\right., \nonumber\\ \\ 
\Pi^{\mathrm{r}}_{D_{2}}&=
\left\{
  \begin{array}{cl}
0&\text{if}\ \ \ |\eta'|<1\\
(1/\chi)\mathrm{Tr}[{Z(x)Z(z)Z^{-1}(x)Z^{-1}(z)}]&\text{if}\ \ \ |\eta'|=1
  \end{array}\right.,
\end{align*}
 (`$\mathrm{r}$' and `$\mathrm{b}$' indicate even and odd bonds, respectively). Throughout the even-Haldane phase we have $\Pi^{\mathrm{r}}_{D_{2}}=-1$ and $\Pi^{\mathrm{b}}_{D_{2}}=1$\cite{IDMRG-note}, and upon the quantum phase transition at $J=J'>0$, these parameters change to $\Pi^{\mathrm{r}}_{D_{2}}=1$ and $\Pi^{\mathrm{b}}_{D_{2}}=-1$. In the ferromagnetic phase, the $D_{2}$ symmetry is broken, which means we have a degeneracy in the ground state.
%%%%%%%%%%%%%%%%%%%%%%%%%%%%%%%%%%%%%%%%%%%%%%%%%%%%
\begin{table}[t]
\caption{
 Inequivalent projective representations of the symmetries of the bond-alternating spin-$1/2$ Heisenberg model.
}
\begin{ruledtabular}
\begin{tabular}{lrrrrrr}
\textrm{Phase}&
\textrm{ $\Pi_{
D_2
}^{\mathrm{r}}$/$\Pi_{
D_2
}^{\mathrm{b}}$}&
\textrm{$\Pi_{T}^{\mathrm{r}}$/$\Pi_{T}^{\mathrm{b}}$}&
\textrm{$\Pi_{P}^{\mathrm{r}}$/$\Pi_{P}^{\mathrm{b}}$}&
\multicolumn{1}{c}{$\theta_{P}\theta'_{P}$}\\
\colrule
even-Haldane &$ -1/1$ & $-1/1$ & $-1/1$ & $-1$ \\
odd-Haldane & $1/-1$ & $1/-1$ & $1/-1$ & $-1$ \\
ferromagnet & $0/0$ & $0/0$ & $1/1$ & $1$ \\
\end{tabular}
\end{ruledtabular}
\label{tab:Phasefactor}
\end{table}
%%%%%%%%%%%%%%%%%%%%%%%%%%%%%%%%%%%%%%%%%%%%%%%%%%%%

A straightforward proof of our results can be obtained by considering the states of Figs.~\ref{fig:MPS}-(b) and \ref{fig:MPS}-(c). These states faithfully represent the odd-Haldane and even-Haldane phases, respectively. If the singlet states on the blue lines in Fig.~\ref{fig:MPS}-(b) are put into the canonical iMPS form, we obtain 
\begin{align}
&\Gamma_{0}=(0 \quad 1), \;\; \Gamma_{1}=(-1 \quad 0), \nonumber \\
&\overline{\Gamma}_{0}=(1 \quad 0)^{T}, \;\; \overline{\Gamma}_{1}=(0 \quad 1)^{T}, \nonumber \\
&\Lambda=\openone/\sqrt{2}, \;\;  \;\; \overline{\Lambda}=\openone. \nonumber
\end{align}
Using Eqs.~(\ref{SUO}) and (\ref{SU2O}) we find $X(x)=\sigma_{x}$, $X(z)=\sigma_{z}$, and $Z(x)=Z(z)=\openone$, which justify the numerical result of Table~\ref{tab:Phasefactor}. The same procedure for the red lines in Fig.~\ref{fig:MPS}-(c) leads to $Z(x)=\sigma_{x}$, $Z(z)=\sigma_{z}$, and $X(x)=X(z)=\openone$, which is again in agreement with the numerical result of Table~\ref{tab:Phasefactor}.

In a similar manner, one can show that to preserve parity and two-site $TI$, the above equations need to be changed to
\begin{equation}
({\Gamma_{i}})^{T}=\theta_{P} N_{\mathrm{b}}^{-1}\overline{\Gamma}_{i}N_{\mathrm{r}},\quad ({\overline{\Gamma}_{i}})^{T}=\theta'_{P} N_{\mathrm{r}}^{-1}\Gamma_{i}N_{\mathrm{b}},   
\end{equation}
where $N_{\mathrm{b}}=\pm N^{T}_{\mathrm{b}}$, $N_{\mathrm{r}}=\pm N^{T}_{\mathrm{r}}$, and $\theta'_{P}\theta_{P}=\pm 1$. Each of these signs ($\pm$) defines a unique label for the phases. Thus, the model with the $P$ and two-site $TI$ symmetries can only show eight phases. Similarly, the parameter of the $P$ symmetry is 
\begin{equation*}
\Pi^{\mathrm{r}(\mathrm{b})}_{P}=
\left\{
  \begin{array}{cl}
0&\text{if}\ \ \ |\eta'|<1\\
(1/\chi)\mathrm{Tr}[{N_{\mathrm{r}(\mathrm{b})}N^{\ast}_{\mathrm{r}(\mathrm{b})}}]&\text{if}\ \ \ |\eta'|=1
  \end{array}.\right.
\end{equation*}
The parameter $\Pi^{\mathrm{r}(\mathrm{b})}_{P}$, in the even-Haldane phase, is $-1$ ($1$), and flips to $1$ ($-1$) for the odd-Haldane phase. The singlet states of Figs.~\ref{fig:MPS}-(b) and \ref{fig:MPS}-(c) lead to $N_{\mathrm{b}}=\sigma_{y}$, $N_{\mathrm{r}}=\openone$, $\theta_{P}\theta_{P}'=-1$, and $N_{\mathrm{r}}=\sigma_{y}$, $N_{\mathrm{b}}=\openone$, $\theta_{P}\theta_{P}'=-1$, respectively, which again justify our numerical results. Note that unlike the $D_2$ and time-reversal symmetries, parity is preserved in the ferromagnetic phase.
%%%%%%%%%%%%%%%%%%%%%%%%%%%%%%%%%%%%%%%%%%%%%%%%%%%%
\begin{table*}
\caption{
Inequivalent projective representations of the symmetries of the most general two-body bond-alternating spin-$1/2$ model with the $[TPD_{2}]$ symmetries [Eq.~\eqref{Eq:GeneralH}].
}
\begin{ruledtabular}
\begin{tabular}{lrrrrrrr}%{cccccccc}
\textrm{Phase}&
\textrm{ $\Pi^{\mathrm{r}}_{D_{2}}$/$\Pi^{\mathrm{b}}_{D_{2}}$ }&
\textrm{ $\Pi^{\mathrm{r}}_{T}$/$\Pi^{\mathrm{b}}_{T}$ }&
\textrm{ $\Pi^{\mathrm{r}}_{P}$/$\Pi^{\mathrm{b}}_{P}$ }&
\textrm{ $\gamma^{\mathrm{r}}_{T}(x)$/$\gamma^{\mathrm{b}}_{T}(x)$ }&
\textrm{ $\gamma^{\mathrm{r}}_{T}(z)$/$\gamma^{\mathrm{b}}_{T}(z)$ }&
\textrm{ $\gamma^{\mathrm{r}}_{P}(x)$/$\gamma^{\mathrm{b}}_{P}(x)$ }&
\textrm{ $\gamma^{\mathrm{r}}_{P}(z)$/$\gamma^{\mathrm{b}}_{P}(z)$}\\
\colrule
Trivial & $1/1$ & $1/1$ & $1/1$ & $1/1$ & $1/1$ & $1/1$ & $1/1$ \\
even-Haldane & $-1/1$ & $-1/1$ & $-1/1$ & $-1/1$ & $-1/1$ & $-1/1$ & $-1/1$ \\
$T_{x}$ &  $1/1$ & $-1/1$ & $-1/1$ & $-1/1$ & $-1/1$ & $-1/1$ & $-1/1$ \\
$T_{y}$ & $1/1$ & $-1/1$ & $-1/1$ & $1/1$ & $-1/1$ & $-1/1$ & $-1/1$ \\
$T_{z}$ & $1/1$ & $-1/1$ & $-1/1$ & $-1/1$ & $1/1$ & $-1/1$ & $-1/1$ \\
$T_{xx}$ & $1/1$ & $-1/1$ & $-1/1$ & $1/1$ & $1/1$ & $-1/1$ & $-1/1$ \\
\end{tabular}
\end{ruledtabular}
\label{tab:T+D_{2}}
\end{table*}
%%%%%%%%%%%%%%%%%%%%%%%%%%%%%%%%%%%%%%%%%%%%%%%%%%%%

To preserve the time-reversal symmetry, we have
\begin{align}
\sum_{i} v_{ji}({\Gamma_{i}})^{\ast}&=M_{\mathrm{r}}^{-1}\Gamma_{j}M_{\mathrm{b}},
\\ \sum_i v_{ji} ({\overline{\Gamma}_{i}})^{\ast}&=M_{\mathrm{b}}^{-1}\overline{\Gamma}_{j}M_{\mathrm{r}},   \label{Time}
\end{align} 
where $vv^{\ast}=-\openone$. Distinct phases are identified by $M_{\mathrm{r}}=\pm M^{T}_{\mathrm{r}}$ and $M_{\mathrm{b}}=\pm M^{T}_{\mathrm{b}}$. 
The parameter associated to this symmetry is
\begin{equation*}
\Pi^{\mathrm{r}(\mathrm{b})}_{T}=
\left\{
  \begin{array}{cl}
0&\text{if}\ \ \ |\eta'|<1\\
(1/\chi)\mathrm{Tr}[{M_{\mathrm{r}(\mathrm{b})}M^{\ast}_{\mathrm{r}(\mathrm{b})}}]&\text{if}\ \ \ |\eta'|=1
  \end{array}.\right.
\end{equation*}
$\Pi^{\mathrm{r}(\mathrm{b})}_{T}$ changes from $-1$ ($1$) in the even-Haldane phase to $1$ ($-1$) in the odd-Haldane phase. One can simply obtain $(M_{\mathrm{r}}=\openone, M_{\mathrm{b}}=i\sigma_{y})$ and $(M_{\mathrm{r}}=i\sigma_{y}, M_{\mathrm{b}}=\openone)$ the for states of Figs.~\ref{fig:MPS}-(b) and \ref{fig:MPS}-(c), respectively. 

Overall, according to Table \ref{tab:Phasefactor}, we assign a label $(\Pi^{\mathrm{r}}_{D_{2}},\Pi^{\mathrm{r}}_{P},\Pi^{\mathrm{r}}_{T}, \Pi^{\mathrm{b}}_{D_{2}}, \Pi^{\mathrm{b}}_{P},\Pi^{\mathrm{b}}_{T} )$ to each phase. Therefore, the labels for even-Haldane, odd-Haldane and ferromagnetic phases are $(-1,-1,-1,1,1,1)$,  $(1,1,1,-1,-1,-1)$, and $(0,1,0,0,1,0)$, respectively---see Fig.~\ref{fig:phase-diagram}. It can be shown that $\Pi^{\mathrm{r}}_{D_{2}}=\Pi^{\mathrm{b}}_{D_{2}}$, $\Pi^{\mathrm{b}}_{P}=\Pi^{\mathrm{r}}_{P}$, and $\Pi^{\mathrm{b}}_{T}= \Pi^{\mathrm{r}}_{T}$ for the Haldane phase of the spin-$1$ Heisenberg chain, which has the one-site $TI$ symmetry recognized by $(-1,-1,-1)$ \cite{Pollmann-2010-Entanglement, Pollmann-2012-Detection}. By comparison of our labels with theses labels, we conclude that the symmetry fractionalization of the red and blue bonds of the even-(odd-)Haldane phase are akin to the Haldane (trivial) and trivial (Haldane) phases, respectively.     

%%%%%%%%%%%%%%%%%%%%%%%%%%%%%%%%%%%%%%%%%%%%%%%%%%%%
\subsection{General bond-alternating model}

We now investigate SPT phases of the most general two-body bond-alternating spin-$1/2$ Hamiltonian, which respects the $[TPD_2]$ symmetries. This general Hamiltonian reads 
\begin{align}
H_{\mathrm{BA}}=&\sum_{i}\sum_{\alpha=1}^{3} \Big(J_{\alpha} \sigma_{\alpha}^{2i}\sigma_{\alpha}^{2i+1} + J'_{\alpha} \sigma_{\alpha}^{2i+1}\sigma_{\alpha}^{2i+2}  \Big),
\label{Eq:GeneralH}
\end{align}
where $\bm{J}=J\big(\sin(\theta)\cos(\varphi), \sin(\theta)\sin(\varphi), \cos(\theta)\big)$ (and similarly for $\bm{J}'$). Within the space of the coupling parameters $(J_{r}\equiv J/J', \theta, \varphi, \theta', \varphi')$ we want to find (numerically) the phases respecting the following conditions: (i) nonvanishing gap, (ii) even entanglement spectrum for one of the bonds, and (iii) respecting the $[TPD_{2}]$ symmetries. We use the procedure of Appendix \ref{App} to find the symmetry fractionalization of the phases of the Hamiltonian (\ref{Eq:GeneralH}). 

A remark is in order here. Results of iDMRG calculations may produce states resembling properties of SPT phases while indeed belonging to a symmetry-breaking class. Fore example, the ground state of the one-dimensional Ising model can be a cat state, that shows even degeneracy in the entanglement spectrum. Thereby, one may erroneously consider the corresponding phase as a SPT phase. To avoid such cases, we employ the following two methods to authenticate the SPT phases: (i) perturbing the SPT phase with proper perturbative terms, and (ii) initializing the iDMRG algorithm with different states. 

To obtain a comprehensive classification of different phases, we need to consider all symmetries together. The combination of the symmetries can produce new labels for the phases, which are specified by the ``commutation relations" between their representations. The combinations of Eqs. (\ref{SUO})-(\ref{Time}) lead to the following commutation relations:
\begin{align}
N_{\mathrm{r}}^{-1}Z(x)N_{\mathrm{r}}&=\gamma_{P}^{\mathrm{r}}(x) Z^{\ast}(x),\label{gamma1}  \\
N_{\mathrm{b}}^{-1}X(x)N_{\mathrm{b}}&=\gamma_{P}^{\mathrm{b}}(x) X^{\ast}(x),   \\
M_{\mathrm{r}}^{-1}Z(x)M_{\mathrm{r}}&=\gamma_{T}^{\mathrm{r}}(x) Z^{\ast}(x), \\
M_{\mathrm{b}}^{-1}X(x)M_{\mathrm{b}}&=\gamma_{T}^{\mathrm{b}}(x) X^{\ast}(x),  
\label{gamma}
\end{align}
which also hold for the $z$ index. Furthermore, if we fix the arbitrary phases of $X(x)$ and $Z(x)$ by imposing $X^2(x)=Z^2(x)=\openone$ and $X^2(z)=Z^2(z)=\openone$, we arrive at the following labels:
\begin{align*}
\gamma(x)&=(\gamma_{P}^{\mathrm{r}}(x), \gamma_{P}^{\mathrm{b}}(x),\gamma_{T}^{\mathrm{r}}(x), \gamma_{T}^{\mathrm{b}}(x)),\\
\gamma(z)&=(\gamma_{P}^{\mathrm{r}}(z)), \gamma_{P}^{\mathrm{b}}(z),\gamma_{T}^{\mathrm{r}}(z), \gamma_{T}^{\mathrm{b}}(z)),
\end{align*}
whose elements could be $\pm1$. These labels are added to the ones obtained previously for a complete classification. In the case of the $T$, $P$, $\mathrm{SU}(2)$, and two-site $TI$ symmetries we obtain $\gamma(x)=\gamma(z)=(1,1,1,1)$, which does not add new phases to the content of  Table~\ref{tab:Phasefactor}. Hence, Table~\ref{tab:Phasefactor} already provide a complete classification of the phases of the bond-alternating spin-$1/2$ Heisenberg model.

We calculate $\gamma(x)$ and $\gamma(z)$ numerically for the most general two-body bond-alternating spin-$1/2$ model
defined in Eq. (\ref{Eq:GeneralH}). In this respect, we sweep the coupling parameter space for $-2\leq J_r \leq 2$ with $\Delta J_r=0.2$, $0\leq \theta, \theta' \leq \pi$ with $\Delta \theta=\Delta \theta'=0.3$, and  $0\leq \varphi, \varphi' \leq 2\pi$ with $\Delta \varphi=\Delta \varphi'=0.5$. Additionally, the iDMRG calculations are performed with $\chi=16$, which lead to the results presented in Table~\ref{tab:T+D_{2}}. Hence, the general Hamiltonian (\ref{Eq:GeneralH}) exhibits symmetry-breaking phases (with degenerate ground states), gapless phases, and twelve different types of SPT phases. Table~\ref{tab:T+D_{2}} shows the symmetry fractionalization of six SPT phases of the model (the other six SPT phases can be obtained by replacing $(\mathrm{r},\mathrm{b})\rightarrow (\mathrm{b},\mathrm{r})$). These SPT phases are labelled with $T$-index similar to the notation of  Ref.~\onlinecite{Liu-2011-Symmetry-protected}. However, we do not claim that Table~\ref{tab:T+D_{2}} necessarily shows all possible SPT phases of the Hamiltonian (\ref{Eq:GeneralH}), because it really depends on the specific values of the coupling parameters. For example, there might still be an SPT phase within a tiny area of the phase diagram, which needs a more fine-tuned coupling parameters and a more careful implementation of iDMRG.    
%%%%%%%%%%%%%%%%%%%%%%%%%%%%%%%%%%%%%%%%%%%%%%%%%%%%
\begin{figure*}[tp]
\vskip5mm
\includegraphics[width=0.66\columnwidth]{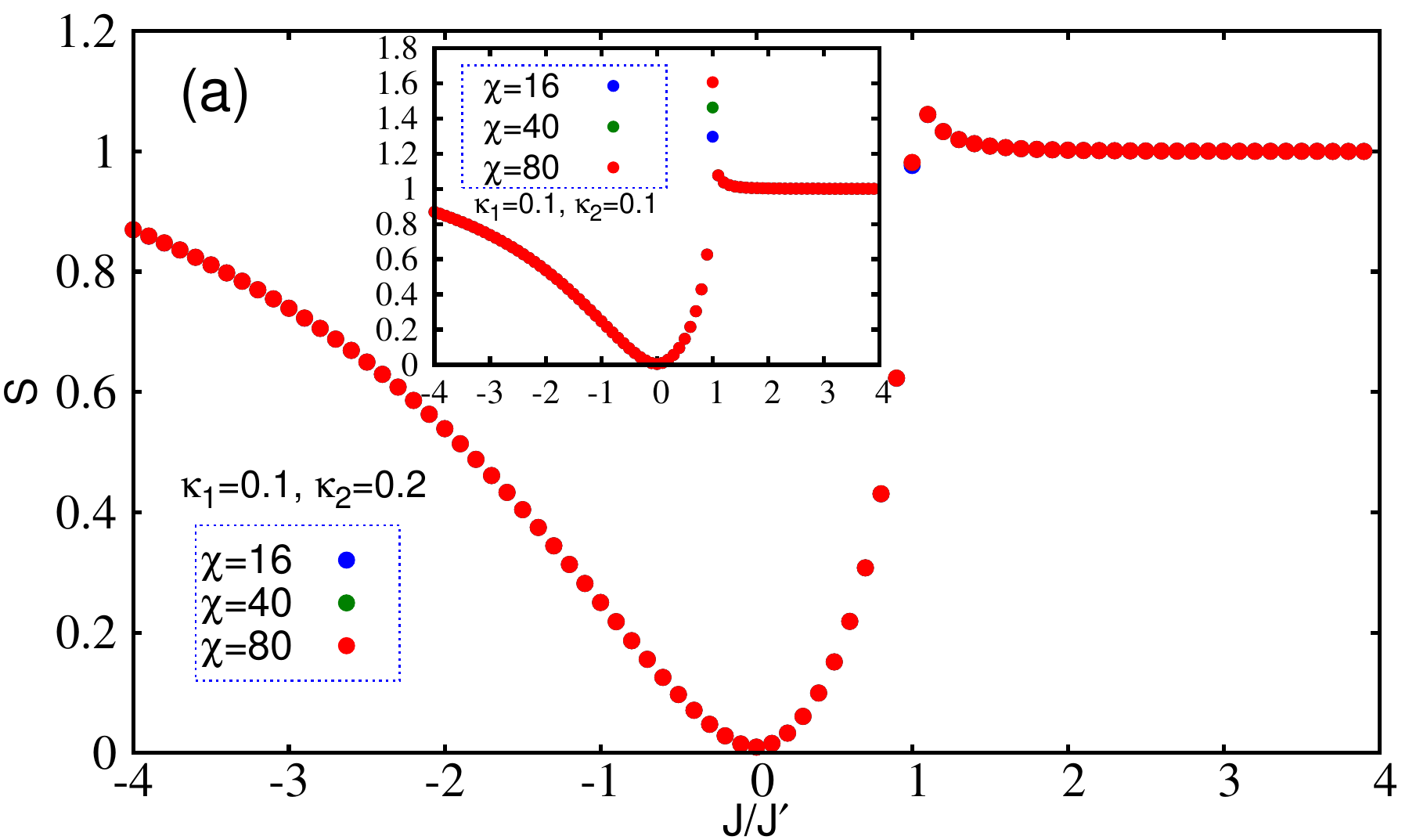}
\hspace{1mm}
\includegraphics[width=0.66\columnwidth]{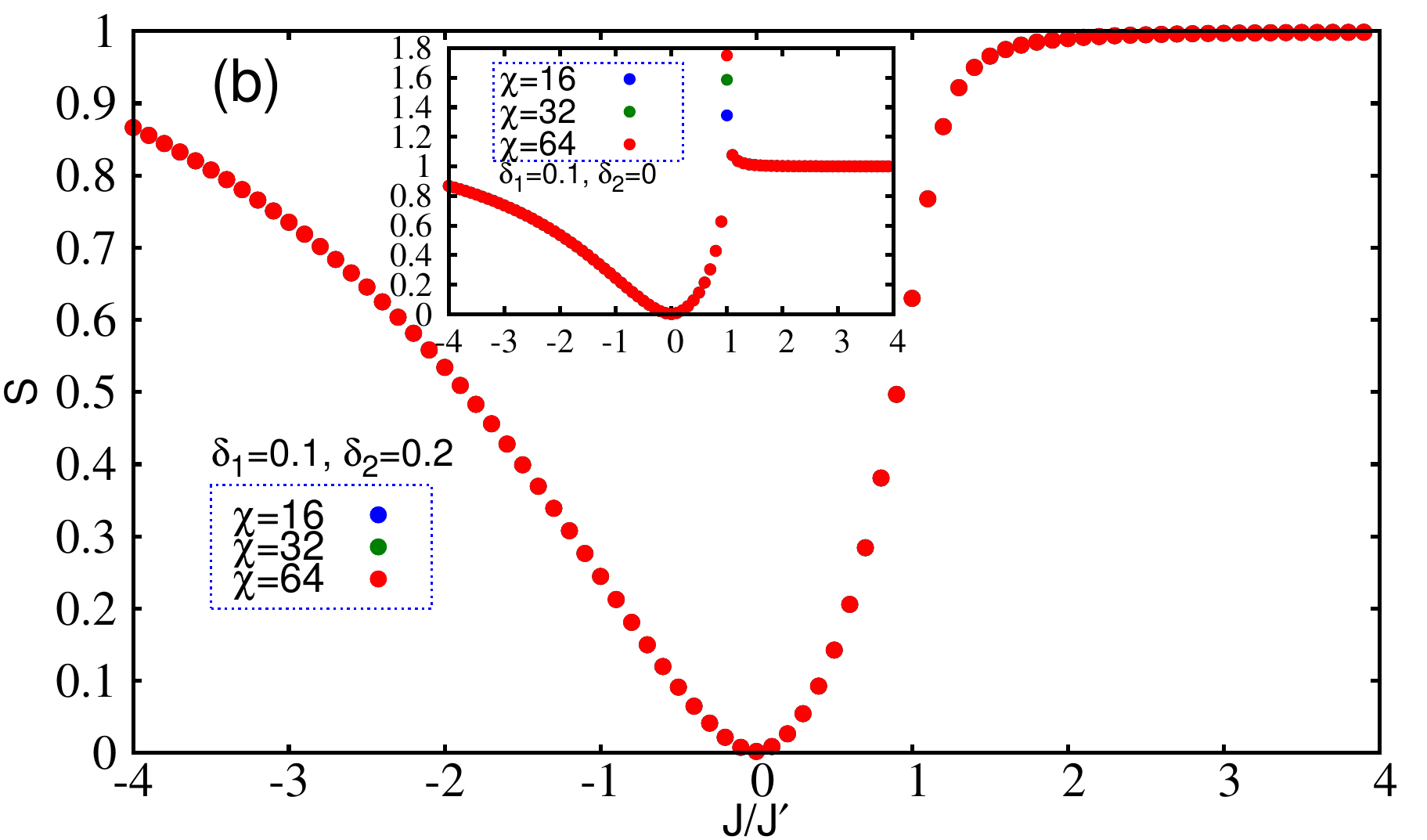}
\hspace{1mm}
\includegraphics[width=0.66\columnwidth]{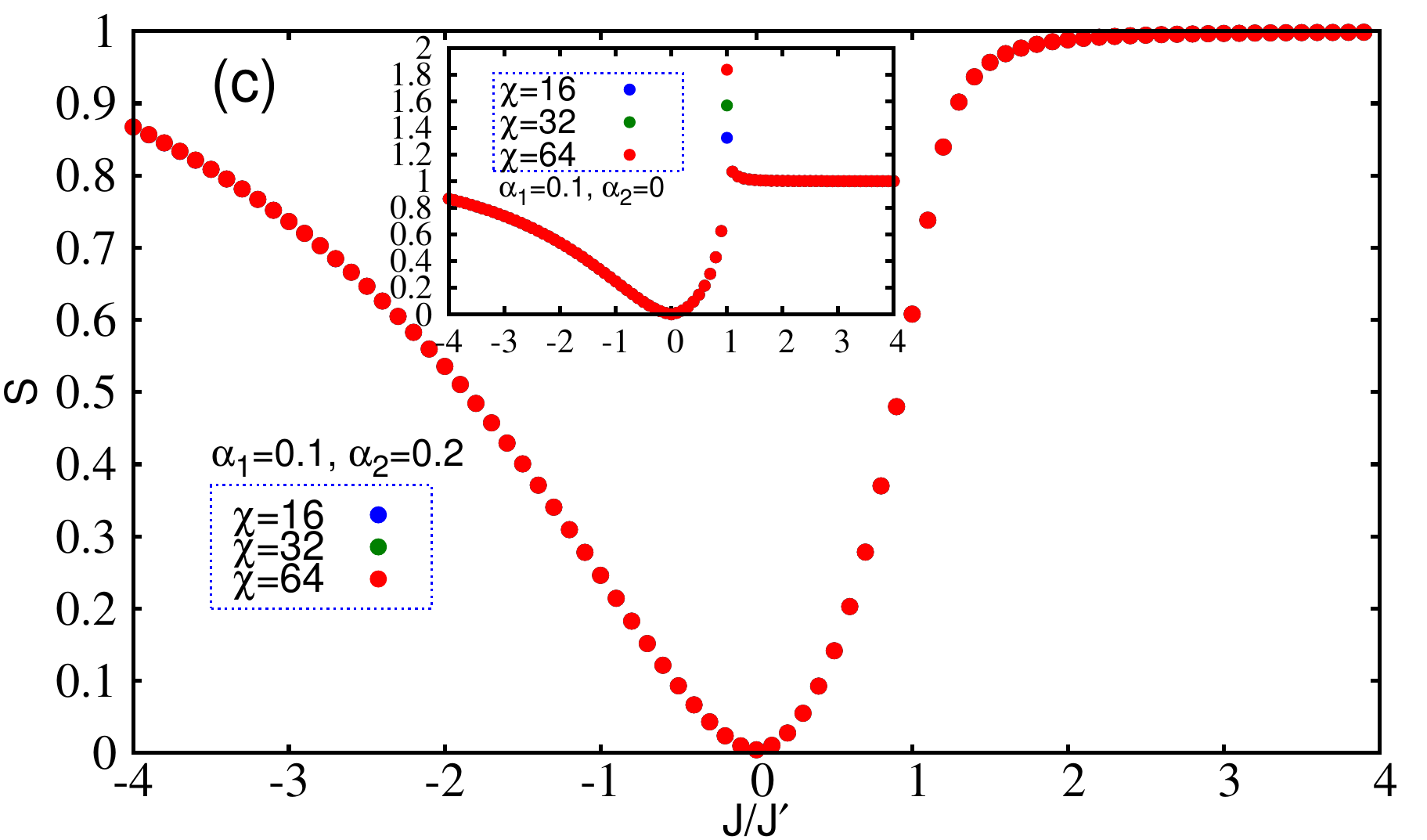}
\caption{(Color online)  The von-Neumann entropy $S$ vs $J$ for different values of $\chi$. (a) The von-Neumann entropy of the Hamiltonian $H+H_{1}$ with $(\kappa_{1}, \kappa_{2})=(0.1, 0.2)$ does not show finite entanglement effects for different $\chi$s, which confirms a single phase in the whole region---specifically, a trivial one. Inset: For $(\kappa_{1},\kappa_{2})=(0.1,0.1)$, the $P$ symmetry is preserved, which results in two different SPT phases separated at the quantum critical point $J=J'$, where it shows a divergent behavior for $S$. (b) The von-Neumann entropy of the Hamiltonian $H+H_{2}$ with $(\delta_{1}, \delta_{2})=(0.1, 0.2)$, which does not show a finite entanglement scaling for different $\chi$s, confirming a single phase. Inset: $\delta_{2}=0$ retrieves the $T$ symmetry and leads to finite entanglement effect at $J=J'$, a signature of quantum phase transition between two SPT phases. (c) $S$ for the Hamiltonian $H+H_3$ and $(\alpha_1, \alpha_2)=(0.1,0.2)$, where the model is in a single phase without phase transition. Inset: For $\alpha_2=0$, which retrieves the $D_2$ symmetry a quantum phase transition between the two SPT phases is observed at $J=J'$.}
\label{fig:VNECluster}
\end{figure*}
%%%%%%%%%%%%%%%%%%%%%%%%%%%%%%%%%%%%%%%%%%%%%%%%%%%%

%%%%%%%%%%%%%%%%%%%%%%%%%%%%%%%%%%%%%%%%%%%%%%%%%%%%
\subsection{Stability of the SPT phases}
\label{subsec:stability}

Here we study the stability of the SPT phases of the bond-alternating spin-$1/2$ Heisenberg model under the breaking of 
the $[TPD_2]$ symmetries. In particular, we would like to clarify whether these symmetries already suffice to protect the SPT 
phases of the model---or perhaps another hidden symmetry is responsible for this. To answer this question, we examine how the addition of three different terms ($H_1$, $H_2$, and $H_3$) to the model may affect the SPT phases. 

%%%%%%%%%%%%%%%%%%%%%%%%%%%%%%%%%%%%%%%%%%%%%%%%%%%%
\subsubsection{$P$ protection}

Let us add the following perturbative cluster-like term to the Hamiltonian (\ref{Hamt}):
\begin{equation}
H_1=\sum_{i}
\kappa_{1} \sigma_{x}^{(2i)} \sigma_{z}^{(2i+1)}\sigma_{x}^{(2i+2)}+\kappa_{2} \sigma_{x}^{(2i+1)} \sigma_{z}^{(2i+2)}\sigma_{x}^{(2i+3)}.
\label{Ham:Cluster} 
\end{equation}
All $[TPD_2]$ symmetries are broken for any nonzero values of $\kappa_{1}\neq \kappa_{2}$, while the $P$ symmetry is retrieved when $\kappa_{1}=\kappa_{2}$. For $\kappa_{1}\neq \kappa_{2}$, $H_{1}$ destroys the even degeneracy of the red (blue) bonds in the even- (odd-) Haldane phase, which verifies that the mentioned symmetries are necessary to protect the even-/odd-Haldane phases. We would like to add that when $\kappa_{1}=\kappa_{2}$, $P$ is preserved, and the even degeneracy of the entanglement spectrum appears immediately. Therefore, $P$ alone can protect the SPT phase. Moreover, adding a cluster term with $\kappa_{1} \neq \kappa_{2}$ prevents the model from exhibiting any quantum phase transition in the whole $J$ region. In fact, the cluster term obstructs the gap closing on the critical line $J=J'$. To show this explicitly, we plot the von Neumann entropy 
\begin{equation}
S= -\sum_{i=1}^{\chi} (\overline{\Lambda})_{ii}\log((\overline{\Lambda})_{ii})
\label{eq:vnentropyj}
\end{equation}
 vs $J$ in Fig.~\ref{fig:VNECluster}-(a). It is expected that entropy diverges as $S \propto \log(\chi)$ \citep{Vidal-2003-Entanglement} whenever a model encounters a quantum phase transition. Figure~\ref{fig:VNECluster}-(a) shows that for $(\kappa_{1},\kappa_2)=(0.1,0.2)$, $S$ remains almost equal for different values of $\chi$, which demonstrates that we encounter a single phase. This is a result of the cluster term, which breaks all necessary symmetries to protect the SPT phases. However, for  $(\kappa_{1}, \kappa_{2})=(0.1,0.1)$, the $P$ symmetry is retrieved, and a finite entanglement effect at $J=J'$ appears, where $S$ shows increasing values for different $\chi$ resembling a divergent like behavior---see the inset of Fig.~\ref{fig:VNECluster}-(a).

%%%%%%%%%%%%%%%%%%%%%%%%%%%%%%%%%%%%%%%%%%%%%%%%%%%%
\subsubsection{$T$ protection}

We examine another perturbation as below to emphasize the protection by $T$ symmetry, 
\begin{eqnarray}
H_{2}=\sum_{i} \big[&&\delta_{1}(\sigma^{(2i)}_{x}\sigma^{(2i+1)}_{z}+\sigma^{(2i+1)}_{x}\sigma^{(2i+2)}_{z}) \nonumber \\
+&&\delta_{2}(\sigma^{(2i)}_{z} +\sigma^{(2i+1)}_{x}) \big].
\label{Ham:TransferField}
\end{eqnarray}
This Hamiltonian breaks the $[TPD_2]$ symmetries given nonzero values for $\delta_{1}$ and $\delta_{2}$, whereas it only respects the $T$ symmetry if $\delta_{2}=0$. For $\delta_{2}\neq 0$, the entanglement spectrum loses the structure of even degeneracy (not shown here), which is a manifestation of no SPT phase. As shown in Fig.~\ref{fig:VNECluster}-(b), $S$ remains 
constant when $\chi$ increases, which denotes that the model does not undergo a quantum phase transition. However, if we choose $\delta_{2}=0$, in which $T$ symmetry is preserved, finite entanglement effects on $S$ appear at  $J=J'$, confirming a quantum phase transition between two SPT phases---see the inset of Fig.~\ref{fig:VNECluster}-(b). 

%%%%%%%%%%%%%%%%%%%%%%%%%%%%%%%%%%%%%%%%%%%%%%%%%%%%
\subsubsection{$D_2$ protection}

The addition of the third Hamiltonian as below indicates the protection by the $D_2$ symmetry,
\begin{eqnarray}
H_3 = &&\alpha_{1} \sum_{i} \big(\sigma_x^{(2i)} \sigma_y^{(2i+1)} \sigma_z^{(2i+2)} 
+\sigma_x^{(2i+1)} \sigma_y^{(2i+2)} \sigma_z^{(2i+3)} \big)
\nonumber \\
&&+\alpha_{2} \sum_{i}\big( \sigma_z^{(2i)}+ \sigma_x^{(2i+1)} \big).
\label{Ham:xyz} 
\end{eqnarray}
Nonzero values for $\alpha_1$ and $\alpha_2$ break the $[TPD_2]$ symmetries, while it keeps only $D_2$ for $\alpha_1\neq 0$ and $\alpha_2=0$. We have plotted the von-Neumann entropy versus $J/J'$ for $(\alpha_1, \alpha_2)=(0.1,0.2)$ in Fig.~\ref{fig:VNECluster}-(c), which does not show a signature of quantum phase transition. In the absence of symmetry, there is no SPT phase, and the model is in a single phase. However, for $\alpha_2=0$ the revival of the $D_2$ symmetry imposes the presence of two SPT phases, which are separated by a quantum phase transition at $J=J'$, as shown in the inset of Fig.~\ref{fig:VNECluster}-(c).

Thus we conclude that any of $P$, $T$ and $D_2$ symmetries protects the even-/odd-Haldane phases on the spin-$1/2$ bond-alternating Heisenberg chain.

%%%%%%%%%%%%%%%%%%%%%%%%%%%%%%%%%%%%%%%%%%%%%%%%%%%%
%%%%%%%%%%%%%%%%%%%%%%%%%%%%%%%%%%%%%%%%%%%%%%%%%%%%
\section{Summary and conclusions}
\label{sec:conc}

We have studied the phases of the bond-alternating spin-$1/2$ Heisenberg chain. Our main tool in doing so has been the recently introduced symmetry fractionalization technique. This technique is based on implementing the known symmetries of the model in the matrix-product representation of the ground state. A set of labels have been obtained from the (inequivalent) 
projective representations of the symmetries (here the time-reversal, parity, and dihedral). These labels can help uniquely identify the phases of the model. 

We have calculated the associated phase labels of this model by employing an infinite-size density-matrix renormalization algorithm and an exhaustive search in the space of Hamiltonian parameters. We identified three phases for this model, one (topologically) trivial phase corresponding to a ferromagnetic state and two symmetry-protected topological phases. We demonstrated that these topological phases naturally resemble a Haldane phase (which originally appeared as the ground state of the spin-$1$ Heisenberg chain or equivalently being in the form of the ground state of the Affleck-Kennedy-Lieb-Tasaki model). As a supporting tool, we also calculated the entanglement spectrum of the model. In addition, by employing the same symmetry fractionalization technique, we also studied the most general one-dimensional bond-alternating model respecting similar symmetries as our model, and found that this model can exhibit twelve different symmetry-protected topological phases. Robustness of the phases of the model against breaking of the time-reversal, parity, and dihedral symmetries have also been investigated. In particular, we perturbed the bond-alternating Hamiltonian with three symmetry-breaking terms in the forms of cluster-like three-body Hamiltonian and two-body interactions of mixed types. Protection of the obtained phases for the model against such perturbations indicated that the set of the symmetries of the system (time-reversal, parity, and dihedral) already suffice to completely characterize the phases.

The bond-alternating spin-$1/2$ Heisenberg chain is a prototype model to demonstrate the spin-Peierls transition and Su-Schrieffer-Heeger model of polyacetylene \cite{SSH:1979}. Hence, our classification in terms of (inequivalent) projective representations of the symmetries is also valid for these models. In other words, the spin-Peierls transition of the spin-$1/2$ Heisenberg chain is the quantum phase transition between two symmetry-protected topological phases, which could be a result of phonon coupling or disorder in the system. Moreover, the one-dimensional representations of the underlying symmetries
could be a classified expression for the Zak phase \cite{Zak:1989}, which has recently been experimentally observed \cite{Atala:2013}.

Our study of complete phase characterization emphasizes the power of the symmetry fractionalization technique for phase identification in one-dimensional gapped systems. We hope that our findings can spur similar investigations on other models of quantum systems. It is evident that developing methods and tools for identification of phases of quantum matter in higher dimensions is certainly an important goal for the next step. 

%%%%%%%%%%%%%%%%%%%%%%%%%%%%%%%%%%%%%%%%%%%%%%%%%%%%
\begin{acknowledgments}

This work was supported partially by Sharif University of Technology's Office of Vice President for Research. A. L. gratefully acknowledges the Alexander von Humboldt Foundation for financial support. We also acknowledge J. Abouie and F. Pollmann for useful discussions and comments. 
\end{acknowledgments}
%%%%%%%%%%%%%%%%%%%%%%%%%%%%%%%%%%%%%%%%%%%%%%%%%%%%

%%%%%%%%%%%%%%%%%%%%%%%%%%%%%%%%%%%%%%%%%%%%%%%%%%%%
\appendix 
 
\section{Numerical calculation of the inequivalent projective representations of a symmetry}
\label{App}

We briefly review how one can employ the iMPS representation of the ground state to completely classify one-dimensional gapped phases. Next, we elaborate on numerical calculations of the inequivalent projective representations of the symmetries of the phases. 
 
\begin{figure}[!h]
\begin{center}
\includegraphics[width=7.0cm,height=9cm]{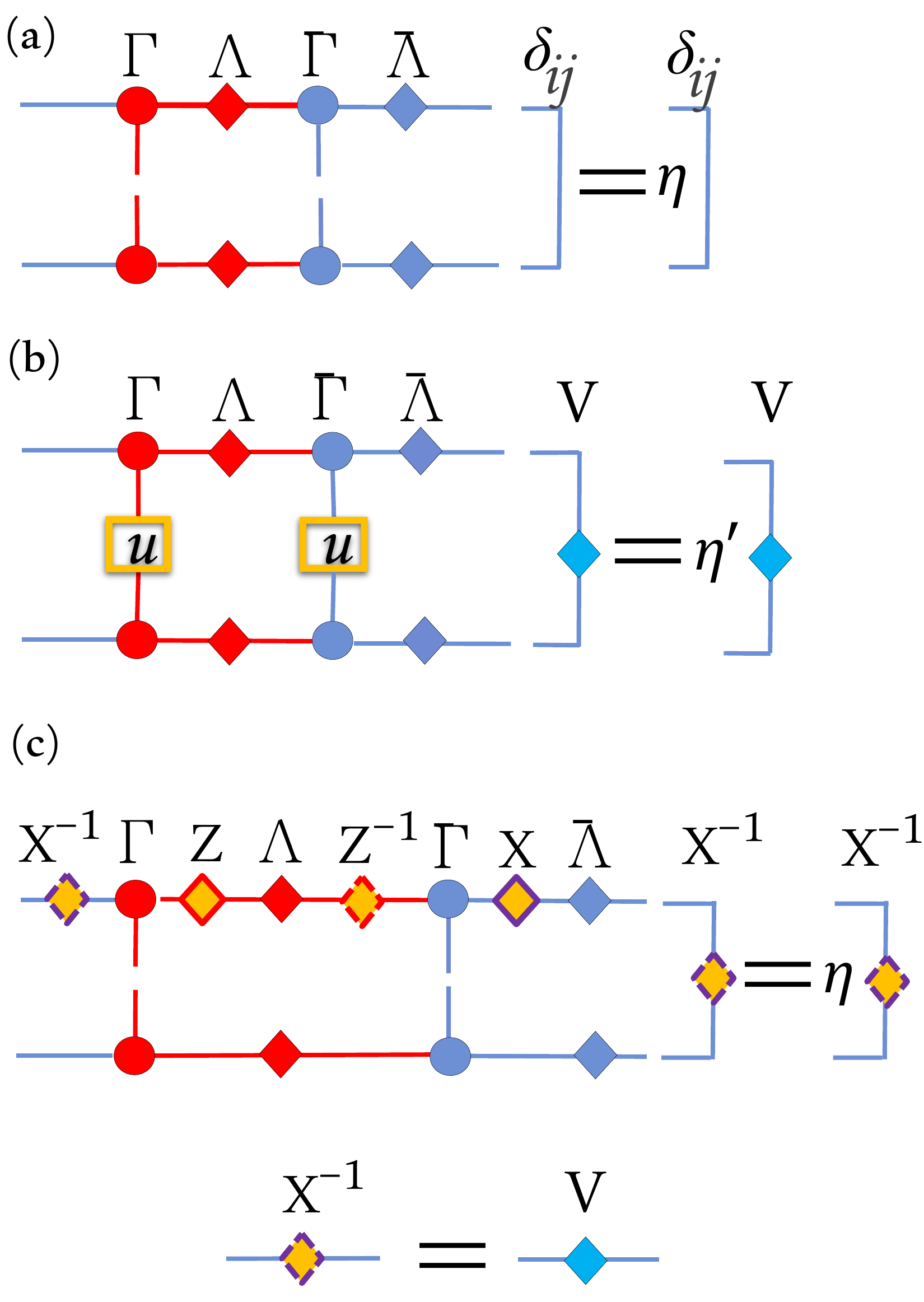}
\caption{(Color online) Schematic procedure for the calculation of inequivalent projective representations of a symmetry. (a) Canonical representation\cite{Orus-2008-Infinite} of short-range correlated iMPS ensures that the two-site transfer matrix $\widehat{T}_{\alpha \beta;ij}\equiv \sum_{mn} (\overline{\Gamma}_{m} \overline{\Lambda} \Gamma_{n} \Lambda)_{\alpha i} (\overline{\Gamma}_{m} \overline{\Lambda} \Gamma_{n} \Lambda)^{\ast}_{\beta j}$ satisfies the following conditions: i. $\sum_{ij}\widehat{T}_{\alpha \beta;ij}\delta_{ij}= \eta \delta_{\alpha \beta}$, and ii. the eigenvalue $\eta$ is non-degenerate and maximum. (b) When a symmetry $u$ acts on iMPS, the two-site transfer matrix changes, $\widehat{T}\rightarrow\overline{T}$. 
Using a large-scale, non-Hermitian eigenvalue solver, such as the Arnoldi\cite{Armadillo} method, one can obtain the eigenvector of $\overline{T}$, i.e., $\overline{T}|V\rangle=\eta'|V\rangle$. (c) If the ground state preserves $u$, Eqs. (\ref{SU}) and (\ref{SU2}) are satisfied. The facts $[\Lambda,X]=0$, $[\Lambda,X^{-1}]=0$, $[\overline{\Lambda},Z^{-1}]=0$, and $[\overline{\Lambda},Z]=0$ ensure that $X^{-1}$ is also an eigenvector of $\overline{T}$. Using the fact that the maximum eigenvalue of $\overline{T}$ is non-degenerate we conclude that $V=X^{-1}$ and $\eta=\eta'$.}
\label{fig:SystemFactor}
\end{center}
\end{figure}

Assume that an iMPS representation is symmetric under the two-site TI and $u(g)^{\otimes N}$, where $u(g)$ is a projective unitary representation of some group $G$. iMPS generally requires two matrices $(\Gamma, \overline{\Gamma})$ to preserve the two-site TI, and preservation of $u(g)^{\otimes N}$ imposes the following condition on $(\Gamma, \overline{\Gamma})$: 
\begin{align}
\sum_{i} u_{ji}(g)\Gamma_{i}&=\beta(g) Z^{-1}(g)\Gamma_{j}X(g), \label{SU}\\
\sum_{i} u_{ji}(g)\overline{\Gamma}_{i}&=\beta'(g) X^{-1}(g)\overline{\Gamma}_{j}Z(g),
\label{SU2}
\end{align} 
where $\beta'(g)$ and $\beta(g)$ are arbitrary phases. The combination of Eqs. (\ref{SU}) and (\ref{SU2}) yields 
\begin{align}
&\sum_{ni} u_{mn}(g) u_{ji}(g)\Gamma_{i}\overline{\Gamma}_{n}=\alpha(g) Z^{-1}(g)\Gamma_{j}\overline{\Gamma}_{m}Z(g),\label{SUreform}\\
&\sum_{ni} u_{mn}(g) u_{ji}(g)\overline{\Gamma}_{i}\Gamma_{n}=\alpha(g) X^{-1}(g)\overline{\Gamma}_{j}\Gamma_{m}X(g),\label{SUreform2}
\end{align} 
where $\alpha(g)=\beta(g)\beta'(g)$. 
Since iMPSs are short-range correlated states (see Fig.~\ref{fig:SystemFactor}-(a)), sufficiently long consecutive sites of $\Gamma_{i}\overline{\Gamma}_{j}$ can span the space of $\chi\times\chi$ matrices. This is called the ``injectivity" property\cite{Garcia-2007-Matrix}. If $u(g)\otimes u(g)$ forms a unitary representation of the group $G'=G\otimes G$, using the injectivity condition, Eqs. (\ref{SUreform}), and (\ref{SUreform2}), we conclude that inequivalent projective representations of $X(g)$ and $Z(g)$ would specify different phases. Moreover, $\alpha(g)$ forms a one-dimensional representation of $G'$, which could be another label for the phases.

%%%%%%%%%%%%%%%%%%%%%%%%%%%%%%%%%%%%%%%%%%%%%%%%%%%%
%%%%%%%%%%%%%%%%%%%%%%%%%%%%%%%%%%%%%%%%%%%%%%%%%%%%

To gain further insight, we first give an example of the $D_{2}$ symmetry. If we denote $u(x)=\sigma_{x}$ and $u(z)=\sigma_{z}$, the representation group of the $D_{2}$ symmetry is defined by $G_{D_{2}}=\langle u(x)\otimes u(x),u(z)\otimes u(z)\rangle$, where $\langle\cdots\rangle$ denotes the generators of the group, which satisfy $u(x)u(z)=u(z)u(x)$ and $u(x)u(x)=u(z)u(z)=\openone$. Equations (\ref{SUreform}) and (\ref{SUreform2}) are written in the following forms:
\begin{align}
&\sum_{ni} u_{mn}(z) u_{ji}(z)\Gamma_{i}\overline{\Gamma}_{n}=\alpha(z) Z^{-1}(z)\Gamma_{j}\overline{\Gamma}_{m}Z(z),\label{D2}\\
&\sum_{ni} u_{mn}(z) u_{ji}(z)\overline{\Gamma}_{i}\Gamma_{n}=\alpha(z) X^{-1}(z)\overline{\Gamma}_{j}\Gamma_{m}X(z),\label{D2-2}
\end{align} 
which also hold for the $x$ index. Using the properties of $G_{D_{2}}$ and the injectivity condition on $\Gamma_{i}\overline{\Gamma}_{j}$, one obtains $\alpha(x)=\pm1$, $\alpha(z)=\pm1$, $X(x)X(z)=\pm X(z)X(x)$, and $Z(x)Z(z)=\pm Z(z)Z(x)$. The first and the second outcomes indicate different one-dimensional representations of the $D_{2}$ symmetry, and the last ones show different projective representations of $D_{2}$. Therefore, four one-dimensional and four projective representation of $D_{2}$ introduce overall sixteen phases.

In the next step, we would like to address the inequivalent projective representations of the two-site $TI$ and $\mathrm{SU}(2)$ symmetries, which is the case of the underlying Hamiltonian in Eq. (\ref{Hamt}). Since $u(g)\otimes u(g)$ for $u(g)\in \mathrm{SU}(2)$ forms a representation group of $\mathrm{SO}(3)$, we only need to consider the $\mathrm{SO}(3)$ symmetry to determine inequivalent projective representations \cite{Haegeman-2012-Order}. Using Eqs. (\ref{SUreform}) and (\ref{SUreform2}), and that $\mathrm{SO}(3)$ has only a one-dimensional representation, we find $\alpha(g)=1$. In contrast to $D_{2}$, one-dimensional representation of $\mathrm{SO}(3)$ does not distinguish the different phases. Thus, different projective representations would
characterize different phases. In fact, $D_{2}$ shows four different projective representation of $\mathrm{SO}(3)$ with the two-site $TI$. Putting all these together, we conclude that the previous equations $X(x)X(z)=\pm X(z)X(x)$ and $Z(x)Z(z)=\pm Z(z)Z(x)$ represent inequivalent projective representations of the two-site $TI$ and $\mathrm{SU}(2)$.

We can obtain a more complete classification of the phases by considering the commutation relation between different symmetries. To clarify this, suppose that the system maintains the $D_{2}$, $P$, $T$, and two-site $TI$ symmetries. To preserve these symmetries, in addition to Eqs. (\ref{D2}) and (\ref{D2-2}), the following equations must be satisfied:
\begin{align}
\sum_{i} v_{ji}({\Gamma_{i}})^{\ast}=&M_{\mathrm{r}}^{-1}\Gamma_{j}M_{\mathrm{b}},\\
\sum_i v_{ji} ({\overline{\Gamma}_{i}})^{\ast}=&M_{\mathrm{b}}^{-1}\overline{\Gamma}_{j}M_{\mathrm{r}}\\
 ({\Gamma_{i}})^{T}=\theta_{P} N_{\mathrm{b}}^{-1}&\overline{\Gamma}_{i}N_{\mathrm{r}},\\
({\overline{\Gamma}_{i}})^{T}=\theta'_{P} N_{\mathrm{r}}^{-1}&\Gamma_{i}N_{\mathrm{b}}. \label{GammaF}
 \end{align}
If we combine Eqs.~(\ref{D2})-(\ref{GammaF}) we find 
\begin{align}
N_{\mathrm{b}}^{-1}X(x)N_{\mathrm{b}}&=\gamma_{P}^{\mathrm{b}}(x) X^{\ast}(x), \label{gamma1}  \\
M_{\mathrm{b}}^{-1}X(x)M_{\mathrm{b}}&=\gamma_{T}^{\mathrm{b}}(x) X^{\ast}(x),  \\
N_{\mathrm{r}}^{-1}Z(x)N_{\mathrm{r}}&=\gamma_{P}^{\mathrm{r}}(x) Z^{\ast}(x),  \\ 
M_{\mathrm{r}}^{-1}Z(x)M_{\mathrm{r}}&=\gamma_{T}^{\mathrm{r}}(x) Z^{\ast}(x), 
\label{gamma}
\end{align}
where $\gamma(x)=(\gamma_{P}^{\mathrm{r}}(x)$, $\gamma_{P}^{\mathrm{b}}(x),\gamma_{T}^{\mathrm{r}}(x)$, and $\gamma_{T}^{\mathrm{b}}(x)$ are some arbitrary phases. Similar relations to Eqs.~(\ref{gamma1})-(\ref{gamma}) are obtained for $z$ instead of $x$. To uniquely define $\gamma(x)$ and $\gamma(z)$, we fix the arbitrary phases of $X(x)$ and $Z(x)$ according to $X^2(x)=Z^2(x)=\openone$ and $X^2(z)=Z^2(z)=\openone$. Under these conditions, $\gamma(x)$ and $\gamma(z)$ can only take $\pm 1$, which is another unique label for the phases \citep{XieChen-2011-Complete}. Therefore, the commutation relation of $D_{2}$, $P$, $T$, and the two-site $TI$ can produce $16\times 16$ different phases. 

To extract numerically the inequivalent projective representation of the underlying symmetries we should obtain a reliable iMPS representation of the ground state to follow the procedure (Fig.~\ref{fig:SystemFactor}). Introducing an appropriate parameter, one can identify inequivalent projective representations of the symmetries and also the commutation relations between them. For instance, the following parameter shows inequivalent projective representation of $D_{2}$:
\begin{align}
\Pi&=
\left\{
  \begin{array}{cl}
0&\text{if}\ \ \ |\eta'|<1\\
(1/\chi)\mathrm{Tr}[{X(x)X(z)X^{-1}(x)X^{-1}(z)}]&\text{if}\ \ \ |\eta'|=1
  \end{array}\right.. \nonumber\\
\end{align}
The commutation relations between $D_{2}$ and $P$ symmetries can be obtained by the following parameter:  
\begin{equation}
\gamma(x)=(1/\chi)\mathrm{Tr}[N^{-1}X(x)NX^{T}(x)], 
\end{equation}
by imposing the constraint $X^2(x)=\openone$.
%%%%%%%%%%%%%%%%%%%%%%%%%%%%%%%%%%%%%%%%%%%%%%%%%%%%

%%%%%%%%%%%%%%%%%%%%%%%%%%%%%%%%%%%%%%%%%%%%%%%%%%%%
\bibliography{BA.bib}

\begin{thebibliography}{38}
\bibitem [{Pro()}]{Projective-note}%
  \BibitemOpen
  \href@noop {} {}\bibinfo {note} {Projective representation of a specific
  symmetry group, e.g., $G$ is accompanied by a phase factor. For instance, if
  elements $g, g'$ and $g''\in G$ have the relation $gg'=g''$, corresponding
  projective representation becomes
  $U_{g}U_{g'}=e^{i\theta}U_{g''}$.}\BibitemShut {Stop}%
\bibitem [{IDM()}]{IDMRG-note}%
  \BibitemOpen
  \href@noop {} {}\bibinfo {note} {To calculate the parameters, we start by
  calculating $\lim_{N \to \infty}\langle \psi|u^{\otimes
  N}(g)|\psi\rangle=\lim_{N \to \infty}(\eta')^{N}$ to see the state possesses
  the symmetry or not. In the SPT phases we find, up to accuracy of machine,
  the ground states preserve the symmetries. After that we calculate some
  quantities such as $\mathrm{Tr}[X(x)X(z)X^{-1}(x)X^{-1}(z)]/\chi$ expected to
  be $\pm 1$. Since the iMPS representation provides a faithful description of
  the gapped systems, the accuracy of calculation of the mentioned quantities
  is again equal to the accuracy of machine. In all parts of our computations
  we use $\chi\leq 80$.}\BibitemShut {Stop}%
\end{thebibliography}
%%%%%%%%%%%%%%%%%%%%%%%%%%%%%%%%%%%%%%%%%%%%%%%%%%%%

\end{document}